\documentclass[conference]{IEEEtran}
\IEEEoverridecommandlockouts

\usepackage{float} 
\usepackage{adjustbox}
\usepackage{cite}
\usepackage{amsmath,amssymb,amsfonts}
\usepackage{algorithmic}
\usepackage{graphicx}
\usepackage{textcomp}
\usepackage{xcolor}
\def\BibTeX{{\rm B\kern-.05em{\sc i\kern-.025em b}\kern-.08em
    T\kern-.1667em\lower.7ex\hbox{E}\kern-.125emX}}

\begin{document}

\title{Low Complexity Autoencoder based End-to-End Learning of Coded Communications Systems \\
\thanks{Authors would like to certify that this work has not been published in any other conference or has not been submitted for any other publication elsewhere.}
}

\author{
\IEEEauthorblockN{Nuwanthika~Rajapaksha, Nandana~Rajatheva,  and Matti Latva-aho}
\IEEEauthorblockA{Centre for Wireless Communications,\\ University of Oulu,\\ Finland\\ 
E-mail: nuwanthika.rajapaksha@oulu.fi, nandana.rajatheva@oulu.fi, matti.latva-aho@oulu.fi}
}

\maketitle

\begin{abstract}
End-to-end learning of a communications system using the deep learning-based autoencoder concept has drawn interest in recent research due to its simplicity, flexibility and its potential of adapting to complex channel models and practical system imperfections. In this paper, we have compared the bit error rate (BER) performance of autoencoder based systems and conventional channel coded systems with convolutional coding (CC), in order to understand the potential of deep learning-based systems as alternatives to conventional systems. From the simulations, autoencoder implementation was observed to have a better BER in 0-5 dB $E_{b}/N_{0}$ range than its equivalent half-rate convolutional coded BPSK with hard decision decoding, and to have only less than 1 dB gap at a BER of $10^{-5}$. Furthermore, we have also proposed a novel low complexity autoencoder architecture to implement end-to-end learning of coded systems in which we have shown better BER performance than the baseline implementation. The newly proposed low complexity autoencoder was capable of achieving a better BER performance than half-rate 16-QAM with hard decision decoding over the full 0-10 dB $E_{b}/N_{0}$ range and a better BER performance than the soft decision decoding in 0-4 dB $E_{b}/N_{0}$ range.

\end{abstract}

\begin{IEEEkeywords}
autoencoder, end-to-end learning, deep learning, neural networks, wireless communications, modulation, channel coding
\end{IEEEkeywords}

\section{Introduction}

Wireless networks and related services have become essential and basic building blocks in the modern society which have changed the life styles to a great extent. Emergence of many new services and applications is challenging the traditional communication landscapes in terms of reliability, latency, energy efficiency, flexibility and connection density, requiring new architectures, algorithms and novel approaches in each and every layer of a communications system.

Communications is a well established field rich of expert knowledge based on information theory, statistics and mathematical modelling. Especially for the physical layer, well proven mathematical approaches for modelling channels \cite{rappaport}, determining optimal signaling and detection schemes for reliable data transmission compensating for hardware imperfections etc. \cite{proakis} are there. However, conventional communications theories display several inherent limitations in achieving the large data and ultra-high-rate communication requirements in complex scenarios. Fast and effective signal processing in latency critical applications, modelling the channels in complex environments, realizing optimum performance in sub-optimal fixed block structured systems are some of the challenges faced by modern communications systems. Applying deep learning concepts to the physical layer has attracted a wider interest in recent history, due to certain advantages of deep learning which could be useful in overcoming above challenges.

Reliable message transmission from a source to destination over a channel with the aid of a transmitter and receiver is the basic requirement of a communications system. In practice, the transmitter and receiver are divided into a series of multiple independent blocks in order to achieve an optimal solution. Even though such a block structure allows individual analysis, controlling and optimization of each block, it is not clear that these individually optimized blocks attain the optimum end-to-end performance. In certain instances, block-based approach is known to be sub-optimal as well \cite{oshea2}. A deep learning-based communications system on the other hand, follows the original definition of a communications system and tries to optimize transmitter and receiver in an end-to-end manner without having an artificial block structure \cite{oshea2}, \cite{oshea1}. Such a straightforward structure which has less energy consumption, lower computational complexity and processing delays seems attracting to be applied in practical systems, especially if such deep learning-based systems can outperform existing conventional systems.

In this study, we have evaluated the BER performance of autoencoder based end-to-end communications systems in additive white Gaussian noise (AWGN) channel, in comparison with the BER performance of equivalent conventional coded communications systems which utilize CC as forward error control mechanism \cite{thesis}\footnote{This paper is based on the research findings from the first author's master's thesis \cite{thesis}.}, \cite{arxiv_paper}\footnote{A pre-conference version of this paper is uploaded to \cite{arxiv_paper}.}. Furthermore, we have implemented a new autoencoder architecture to reduce the model dimensionality and training complexity caused by larger message sizes. The newly proposed autoencoder layout has lower model dimensions compared to the original autoencoder layout proposed in \cite{oshea2} and hence has a lesser number of learnable parameters, thus having a significantly lower training complexity than the originally proposed autoencoder layout. Lower model dimensions of the new autoencoder has also resulted in having lower processing complexity, hence low latency processing compared to the original one. Simulation results have shown that the autoencoder based implementations have comparable BER to the equivalent conventional implementations with channel coding and higher modulation orders such as 16-QAM, showing the potential of autoencoder based end-to-end communications as an alternative to conventional block based communications. 

The structure of the paper is as follows. The following sub section discusses some of the related work in this domain. Section II explains the autoencoder concept and autoencoder based systems as alternatives for coded systems with BPSK modulation and higher order modulations. Section III presents the obtained results for the implemented autoencoder based systems. Section IV concludes the paper.

\subsection{Related Work}

O'Shea et al. first introduced using the autoencoder concept in communications systems \cite{oshea2}, \cite{oshea1}. A novel way of communications system design based on autoencoder is presented in \cite{oshea2}, where the communication task is considered as an end-to-end reconstruction task which jointly optimizes the transmitter and receiver in a single process. The concept of Radio Transformer Networks (RTNs) is presented as a method to incorporate expert domain knowledge in the machine learning model. Extension of the autoencoder model to multiple transmitter and receiver pairs is also presented. In \cite{mimo1}, authors have introduced a novel physical layer scheme for the multiple input multiple output (MIMO) communications, extending the autoencoder based end-to-end learning approach to multi-antenna case. Autoencoder based communications for the single input single output (SISO) and MIMO interference channel in flat-fading conditions is introduced in \cite{mimo2}.

In \cite{oshea3}, the channel autoencoder model is improved enabling end-to-end learning in instances where the channel response is unknown or difficult to model in a closed form analytical solution. By adopting an adversarial approach for channel response approximation and information encoding, jointly optimum solution is learned for both tasks over a wide range of channel environments. They have presented the results of the proposed system with training and validation done for an over-the-air system. In \cite{hoydis}, an autoencoder based over-the-air transmission system is presented where they have built, trained and executed a communications system fully composed of neural networks (NNs) using unsynchronized off-the-shelf software-defined radios. They have also introduced mechanisms for continuous data transmission and receiver synchronization, proving the feasibility of over-the-air implementation of a fully deep learning-based communications system.

\begin{figure}[ht]
\centerline{\includegraphics[width=0.5\textwidth]{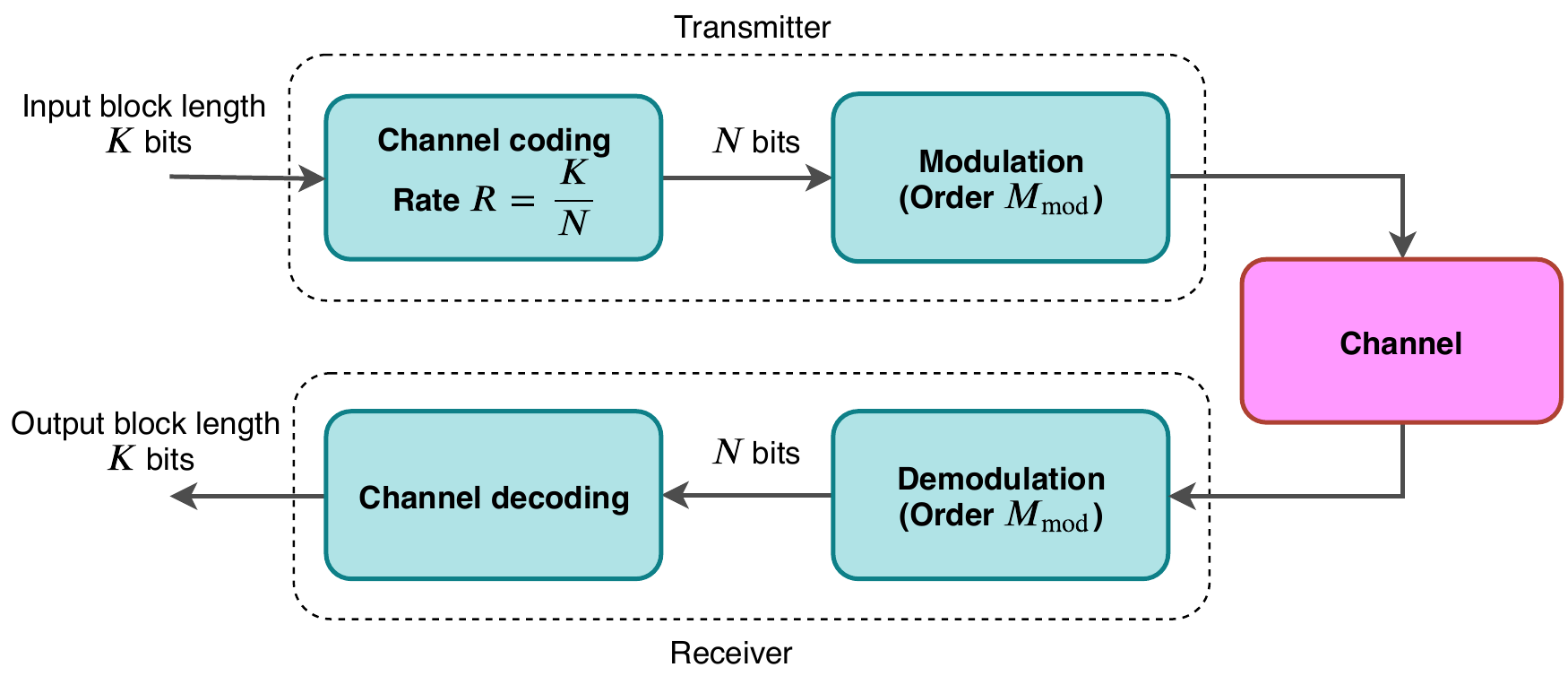}}
\caption{System model of a conventional communications system.}
\label{fig:coded_comsystem}
\end{figure}

More recently, the concept of end-to-end learning is also being applied in optical communications and molecular communications domains, where autoencoder based frameworks are proposed and implemented with comparable performance, showing the potential of deep learning-based end-to-end communications in complex channel conditions and operating environments \cite{optical}, \cite{molecular}.

However, according to our best knowledge, autoencoder performance comparison with respect to advanced channel coded systems and higher order modulation schemes has not been covered in the existing literature so far. In order to address this gap, in this paper, we have investigated the performance of the autoencoder based end-to-end learning, comparing it against conventional communications systems which use CC. Also, the novel autoencoder layout which we have proposed in the paper is shown to have better performance compared to convolutional coded systems with higher order modulations.

\section{Autoencoder based End-to-End Learning of Coded Systems}

As shown in Fig. \ref{fig:coded_comsystem}, a standard communications system consists of several blocks. Channel encoding block with code rate $R$ converts the information block of size $K$ bits to a block of size $N$. Modulator of order $M_{mod}$ then converts $k_{mod}=log_{2}(M_{mod})$ bits into transmit symbols according to the modulation scheme. At demodulation/detection and channel decoding blocks at the receiver, the reverse process is performed in order to recover the estimated information block. 

An autoencoder is a type of artificial neural network that tries to reconstruct its input at the output in an unsupervised manner \cite{goodfellow}. When a communications system is interpreted as an autoencoder based system, instead of having an explicit block structure as in a conventional communications system, autoencoder tries to optimize the overall system in an end-to-end manner. Autoencoder models are implemented equivalent to conventional communications systems based on several system parameters such as the input message size, number of channel uses per message and signal power constraints, enabling performance comparison of the two approaches.

\subsection{End-to-End Learning of Coded Systems with BPSK}
\label{sec:coded_bpsk}

Initially, we have done a BER performance evaluation of autoencoder based system in comparison to conventional channel coded communications system with BPSK modulation over the AWGN channel. For that, we have implemented a model similar to the original autoencoder proposed by \cite{oshea2} with slight variations. Given the task of communicating one out of $M$ possible messages $s\in{\mathbb{M}} = \{1,2,...,M\}$ using $n$ complex channel uses with a minimum reconstruction error, the autoencoder model is a feedforward NN constructed by sequentially combining the layers \{\textit{Input, Dense-ReLU, Dense-ReLU, Dense-Linear, Normalization, Noise, Dense-ReLU, Dense-ReLU, Dense-Softmax}\} which have $\{M,M,M,2n,2n,2n,M,M,M\}$ output dimensions respectively. Layers 1-5 compose the transmitter side of the system where the energy constraint of the transmit signals is guaranteed by the normalization layer at the end. Layers 7-9 compose the receiver side of the system where estimated message can be obtained from the output of the \textit{Softmax} layer. For model training, \textit{Noise} layer in-between the transmitter and receiver side of the model acts as the AWGN channel. The model is trained end-to-end using stochastic gradient descent (SGD) using the categorical cross-entropy loss function on the set of all possible messages $s\in{\mathbb{M}}$.

However, as also pointed out in \cite{oshea2}, it is evident that representation of message $s$ by an $M$-dimensional vector becomes impractical for large $M$ values due to huge memory and processing requirements. Model dimensions significantly increase with $M$ and large number of learnable parameters make it difficult to efficiently train the model. In order to overcome this challenge we have developed a new autoencoder layout which uses a binary vector to represent message $s$ which has only $log_2(M)$ dimensions. Details of the implementation and its performance evaluation are given in the following sections.

\subsection{Low Complexity End-to-End Learning of Coded Systems with Higher Order Modulations}
\label{sec:coded_higher}

This section presents the design and development of the new autoencoder based end-to-end communications system which has a low training and processing complexity compared to the previous autoencoder model. In addition to having binary inputs to reduce the training complexity, we have also improved the original autoencoder layout proposed by \cite{oshea2}, changing the layer dimensions of the model in order to absorb different parameter settings in a conventional communications system. Internal layers of the proposed autoencoder architecture are designed with parameters relating to the channel encoder, modulator functions at the transmitter side and demodulator, channel decoder functions at the receiver side.

\begin{figure*}[ht]
\centerline{\includegraphics[width=1\textwidth]{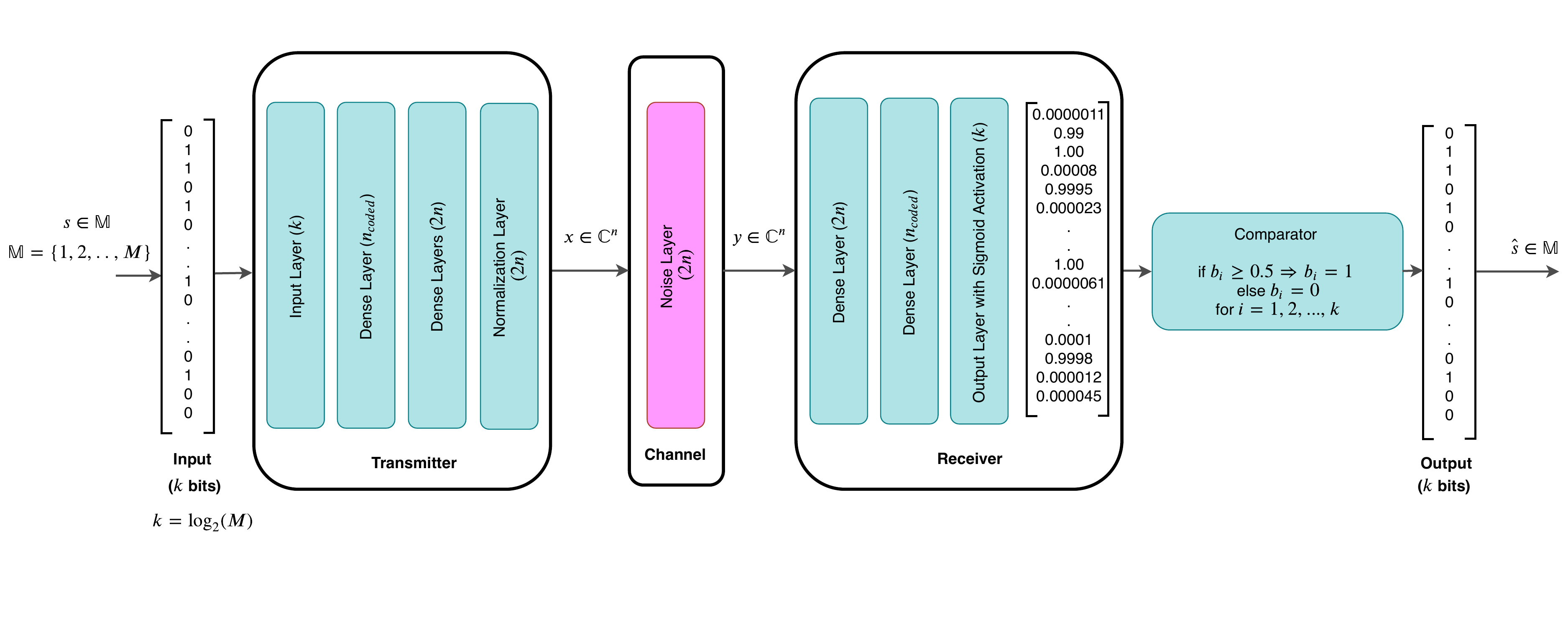}}
\caption{New autoencoder implementation with lower training and processing complexity.}
\label{fig:binary_autoencoder}
\end{figure*}

In the model, the message $s$ is taken in as a binary vector of $k$ bits ($k=log_{2}(M)$). For a code rate of $R$, the number of coded bits per each message is $n_{coded} = k/R$. Then, $n=n_{coded}/k_{mod}$ symbols are required to transmit the encoded $n_{coded}$ bits where $k_{mod}=log_{2}(M_{mod})$ and $M_{mod}$ is the given modulation order. The $K$ bit information blocks used in conventional system are divided into $k$ bits long messages when feeding into the autoencoder. Output of the decoder is of size $k$ bits and gets mapped to the estimated message $\hat{s}$. These parameters are used to determine the dimensions of each layer of the autoencoder model as shown in Table \ref{table:binary_autoencoder}. Fig.~\ref{fig:binary_autoencoder} illustrates the layout of the model.

\begin{table}[ht]
\caption{Layout of the newly proposed low complexity autoencoder}
\begin{center}
\begin{tabular}{|c||c|} 
\hline
Layer & Output dimensions \\    
\hline      
Input & $k$  \\
\hline
Dense-ReLU & $n_{coded}$ \\
\hline
Dense-ReLU & $2n$ \\
\hline
Dense-ReLU & $2n$ \\
\hline
Dense-Linear & $2n$ \\
\hline
Normalization & $2n$ \\
\hline
Noise & $2n$ \\
\hline
Dense-ReLU & $2n$ \\
\hline
Dense-Linear & $n_{coded}$ \\
\hline
Dense-Sigmoid & $k$ \\
\hline
\end{tabular}
\label{table:binary_autoencoder}
\end{center}
\end{table}

A \textit{Sigmoid} layer is used as the output layer along with binary-cross entropy loss function since we deal with binary inputs and outputs. After the model training, autoencoder output is applied to a comparator module which returns the binary outputs as shown in Fig. \ref{fig:binary_autoencoder}. More details about selecting the model layout, layer types and activation functions are available in \cite{thesis}. For model training, \textit{Noise} layer in-between the transmitter and receiver side of the model acts as the AWGN channel.

\section{Simulations and Results}

\subsection{Coded Systems with BPSK Modulation}
\label{sec:bpsk_results}

Autoencoder BER performances were evaluated and compared with their equivalent conventional channel coded BPSK systems using the autoencoder model described in Section~\ref{sec:coded_bpsk}. Different rates $R=log_2(M)/n$ were obtained by changing the message size $M = \{2,4,16,256\}$ and the number of $n$ channel uses, resulting in autoencoder models equivalent to baseline systems with BPSK modulation ($M_{mod}=2$) and code rates $R=\{1/2, 1/3, 1/4 \}$. Noise variance of the AWGN channel is $\beta = (2RE_{b}/N_{0})^{-1}$. 

Model implementation, training and testing was done in Keras~\cite{keras} with TensorFlow~\cite{tensorflow} backend. End-to-end model training over the stochastic channel model was done using SGD with Adam optimizer with a learning rate of 0.001. Equal message transmit energy constraints were kept in each autoencoder model and its corresponding baseline system during simulations for fair comparison. For each model, 1,000,000 random messages were used as the training set and model training was done over 50 epochs with batch size of 2000. $E_{b}/N_{0}$ = 5 dB was used for model training. Model testing was performed with 1,000,000 different random messages over 0 dB to 10 dB $E_{b}/N_{0}$ range. BER performances of the autoencoder models and corresponding baselines were compared for each configuration setting.

Convolutional codes with Viterbi decoder with hard and soft decision decoding was used in the channel encoder/decoder blocks in the baseline system. Information block length for baseline system is taken as $K=800$ and CC with constrain length 7 is used.

\begin{figure}[ht]
\centerline{\includegraphics[width=0.51\textwidth]{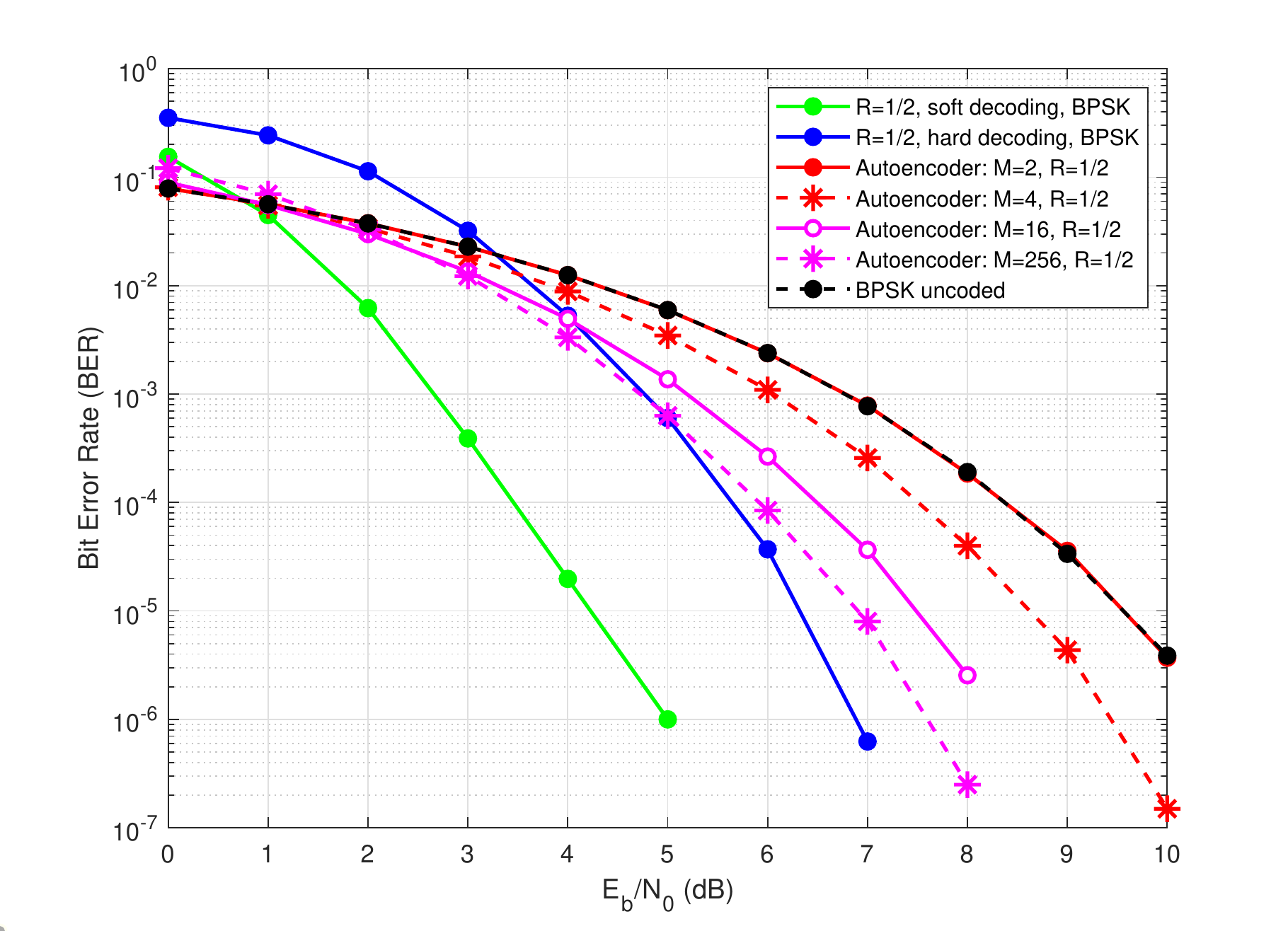}}
\caption{$R=1/2$ system BER performance comparison of different autoencoder models with $M=\{2,4,16,256\}$.}
\label{fig:ber0.5_models}
\end{figure}

\begin{figure}[ht]
\centerline{\includegraphics[width=0.48\textwidth]{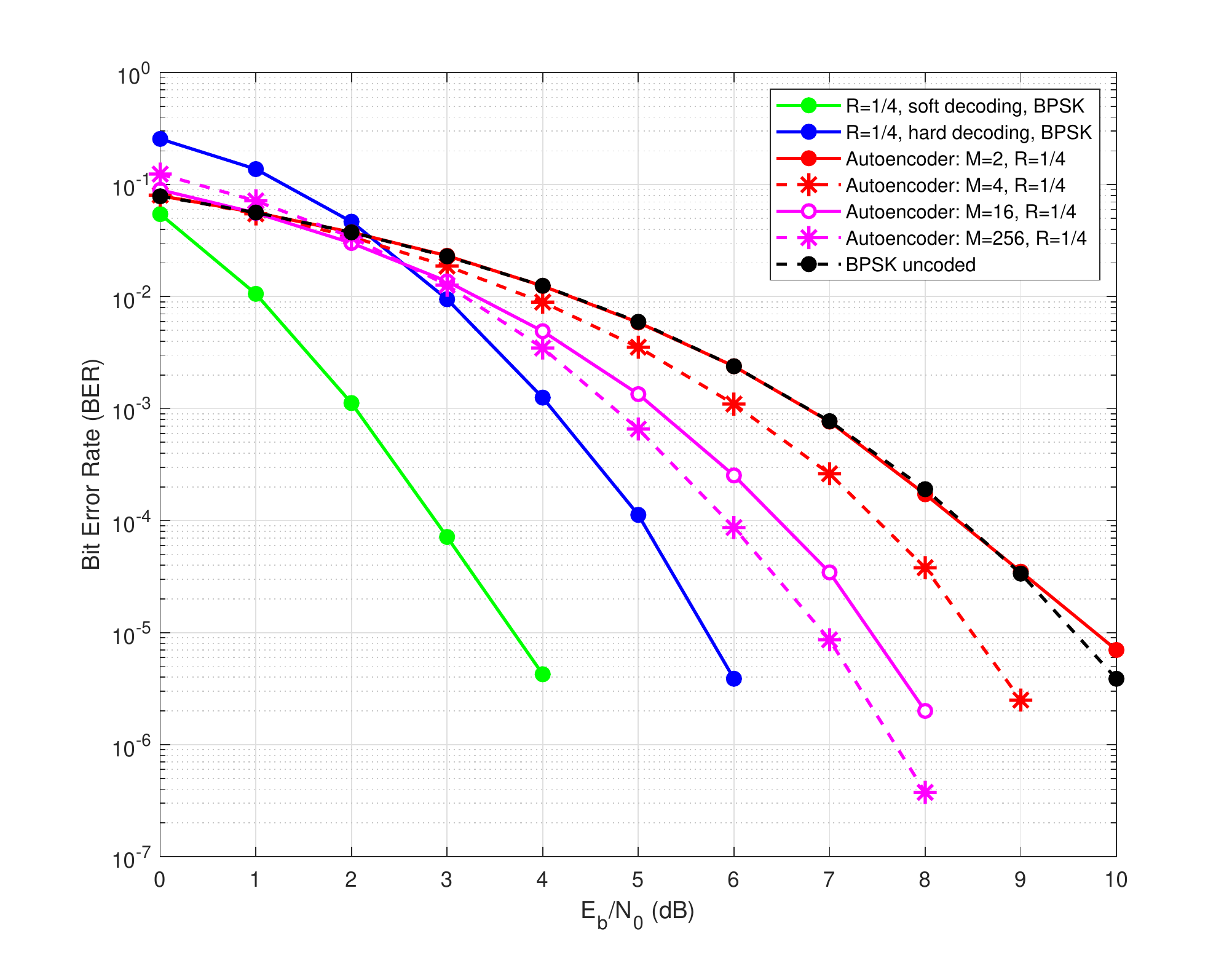}}
\caption{$R=1/4$ system BER performance comparison of different autoencoder models with $M=\{2,4,16,256\}$.}
\label{fig:ber0.25_models}
\end{figure}

BER plots in fig.~\ref{fig:ber0.5_models} and Fig.~\ref{fig:ber0.25_models}, show that the autoencoder BER performance improves with increasing message size. In each figure, BER curve of $M=2$ model overlaps with that of uncoded BPSK where as $M=256$ model has much improved BER curve. Having more degrees of freedom and flexibility in the model due to the increased message size which helps a better end-to-end optimization is the reason for this improvement. $M$ and $R$ values determine the dimensions of the autoencoder model. Not much coding gain is achieved for low $M$ values since the layer dimensions limit the non-linearities introduced by the model during the learning process. Layer dimensions increase with $M$, and for same code rate $R$ model has higher degrees of freedom with more learnable parameters that can be optimized to minimize the end-to-end reconstruction error, enabling it to learn transmit symbols with a coding gain. Number of learnable parameters for two different models are compared in Table \ref{table:model_parameters} explaining how message size impacts the model learning capacity.

From the BER plots, we can observe that autoencoder has a comparable BER performance to the hard decision CC. In $R=1/2$ system, $M=256$ autoencoder outperforms hard decision CC in 0 dB to 5 dB $E_{b}/N_{0}$ range and it is only around 1 dB worse at $10^{-5}$ BER. It should be noted that in each simulation setting, a single model trained at $E_{b}/N_{0}$ = 5 dB gives this BER performance over the full the $E_{b}/N_{0}$ range.

\begin{table}[h]
\caption{Learnable parameters of $R=1/2$ autoencoder models}
\begin{center}
\begin{tabular}{|c||c|c|c|}
\hline
Layer  & \multicolumn{3}{c|}{Number of parameters} \\  \cline{2-4}
(Output dimensions)  & (M, n) &  (M=2, n=2) & (M=256, n=16) \\
\hline
Input (M) & 0  & 0 & 0\\
\hline
Dense-ReLU (M) &(M+1)*M & 6 & 65792\\
\hline
Dense-ReLU (M) & (M+1)*M & 6 & 65792\\
\hline
Dense-Linear (2n) & (M+1)*2n & 12 & 8224\\
\hline
Normalization (2n) & 0 & 0 & 0\\
\hline
Noise (2n)& 0 & 0 & 0\\
\hline
Dense-ReLU (M) & (2n+1)*M & 10 & 8448\\
\hline
Dense-ReLU (M)& (M+1)*M & 6 & 65792\\
\hline
Dense-Softmax (M) & (M+1)*M & 6 & 65792\\
\hline
\end{tabular}
\label{table:model_parameters}
\end{center}
\end{table}

\begin{table}[ht]
\caption{System parameters for autoencoder and baseline systems}
\begin{center}
\begin{tabular}{|c|c|c|c||c|c|}
\hline
\multicolumn{4}{|c||} {Autoencoder configurations}  & \multicolumn{2}{c|} {Baseline system parameters}   \\  \hline
 $M$ & $k=log_2(M)$ & $n_{coded}$ & $n$ & $R$ &  $M_{mod}$ \\
\hline
16 & 4 & 8 & 4 &  1/2 & 4   \\
64 & 6 & 12 & 6 & & (QPSK) \\
256 & 8 & 16 & 8 & &  \\  \hline
256 & 8 & 16 & 4 &  1/2 & 16 \\    
4096 & 12 & 24 & 6 & & (16-QAM)   \\
\hline
\end{tabular}
\label{tab1e:model_configurations_binaryAE}
\end{center}
\end{table}

\subsection{Coded Systems with Higher Order Modulations}

AWGN channel BER performance of the newly proposed autoencoder described in Section~\ref{sec:coded_higher} was evaluated in comparison with conventional channel coded systems with higher order modulations. Different system parameters used for autoencoder implementation are summarized in Table \ref{tab1e:model_configurations_binaryAE}. Noise variance of the AWGN channel is $\beta = (2R k_{mod}E_{b}/N_{0})^{-1}$. All the other training and testing parameters are same as in Section \ref{sec:bpsk_results} other than the batch size being 1000 and the number of epochs being 100.

Fig. \ref{fig:ber_M256,Rc05,Mmod16} shows the BER comparison between the autoencoder and baseline for $R=1/2$ and $M_{mod}=16$ configuration. Autoencoder outperforms hard decision CC across the full $E_{b}/N_{0}$ range considered. It is interesting to see that the autoencoder is better than both soft decision CC and uncoded 16-QAM in low $E_{b}/N_{0}$ range from 0 dB to 4 dB, where soft decision CC has an inferior BER performance than the uncoded 16-QAM. Here also, a single autoencoder model trained at $E_{b}/N_{0}$ = 5 dB gives this BER performance over the full the $E_{b}/N_{0}$ range. Thus, training the model at a given $E_{b}/N_{0}$ has been capable of learning transmit mechanisms across the full $E_{b}/N_{0}$ range showing the learning capability and adaptability of the deep learning model.

\begin{figure}[ht]
\centerline{\includegraphics[width=0.51\textwidth]{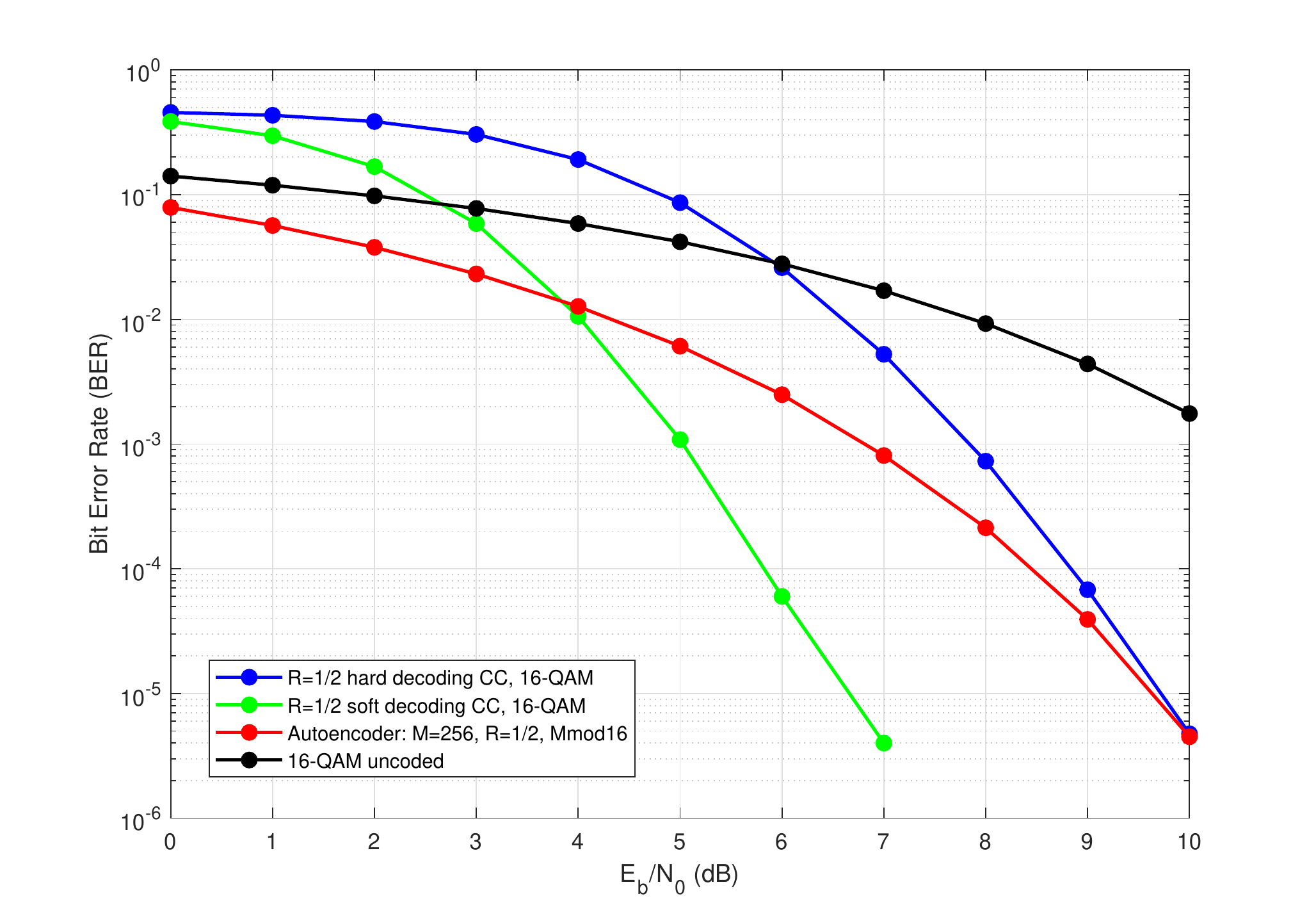}}
\caption{Autoencoder and baseline BER for $R=1/2$ and $M_{mod}=16$.}
\label{fig:ber_M256,Rc05,Mmod16}
\end{figure}

\begin{figure}[ht]
\centerline{\includegraphics[width=0.48\textwidth]{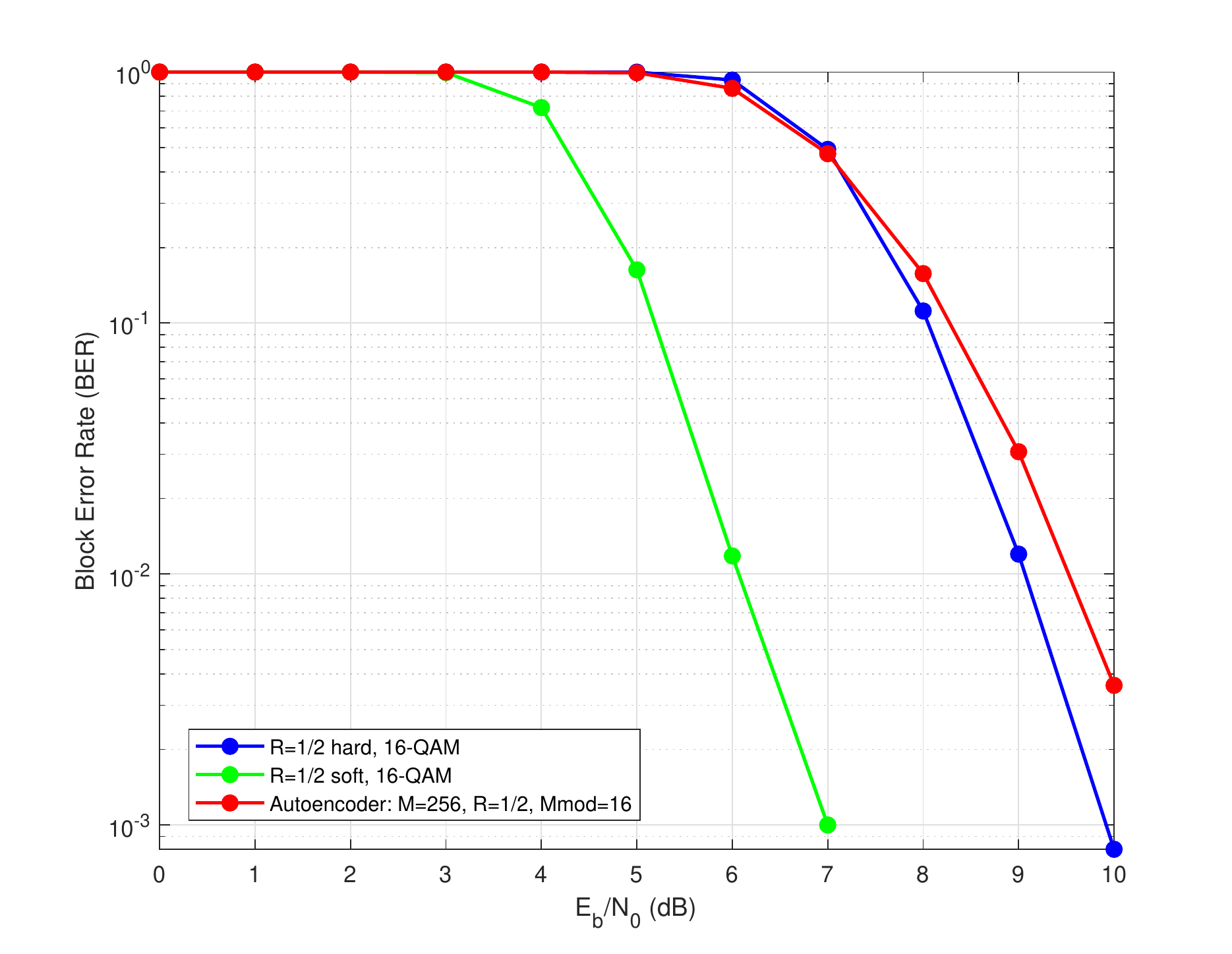}}
\caption{Autoencoder and baseline BLER for $R=1/2$ and $M_{mod}=16$.}
\label{fig:bler_M256,Rc05,Mmod16}
\end{figure}

Fig. \ref{fig:bler_M256,Rc05,Mmod16} shows the block error rate (BLER) comparison between the autoencoder and baseline. Autoencoder has an inferior BLER than the baseline which can be explained as the autoencoder is optimized for a minimum message error rate (MER) or BER of each transmit message instead of a minimum BLER. When considering the MER, autoencoder has an acceptable MER performance as shown in Fig. \ref{fig:mer_M256,Rc05,Mmod16}, having less than $10^{-4}$ error at 10 dB. 

An interesting observation is that the autoencoder achieves acceptable BER and MER with very short input message lengths ($k=8$ in this case) which is a very low block size compared to the block sizes used in existing communications systems which are 100s or 1000s bits long. Such a system having an acceptable error performance with a lower processing complexity and processing delay compared to conventional systems would be advantageous for low latency, low throughput communications where short message transmission is recommended.

\subsection{Training and Processing Complexity}

First we compare the model training complexity of the two autoencoder architectures to understand the efficiency of the proposed autoencoder layout. For a model with message size $M=256$, code rate $R=1/2$ and modulation order $M_{mod}=16$, original autoencoder has 267,528 learnable parameters and input dimensionality of the model is 256. Where as in the new autoencoder, only 776 learnable parameters are there and input vector dimension is only 8 bits. During our simulations, both these models were observed to have a similar BER performance. Therefore we can state that the latter model is more efficient to implement, especially for systems with higher modulation orders where larger $M$ values are required.

\begin{figure}[ht]
\centerline{\includegraphics[width=0.49\textwidth]{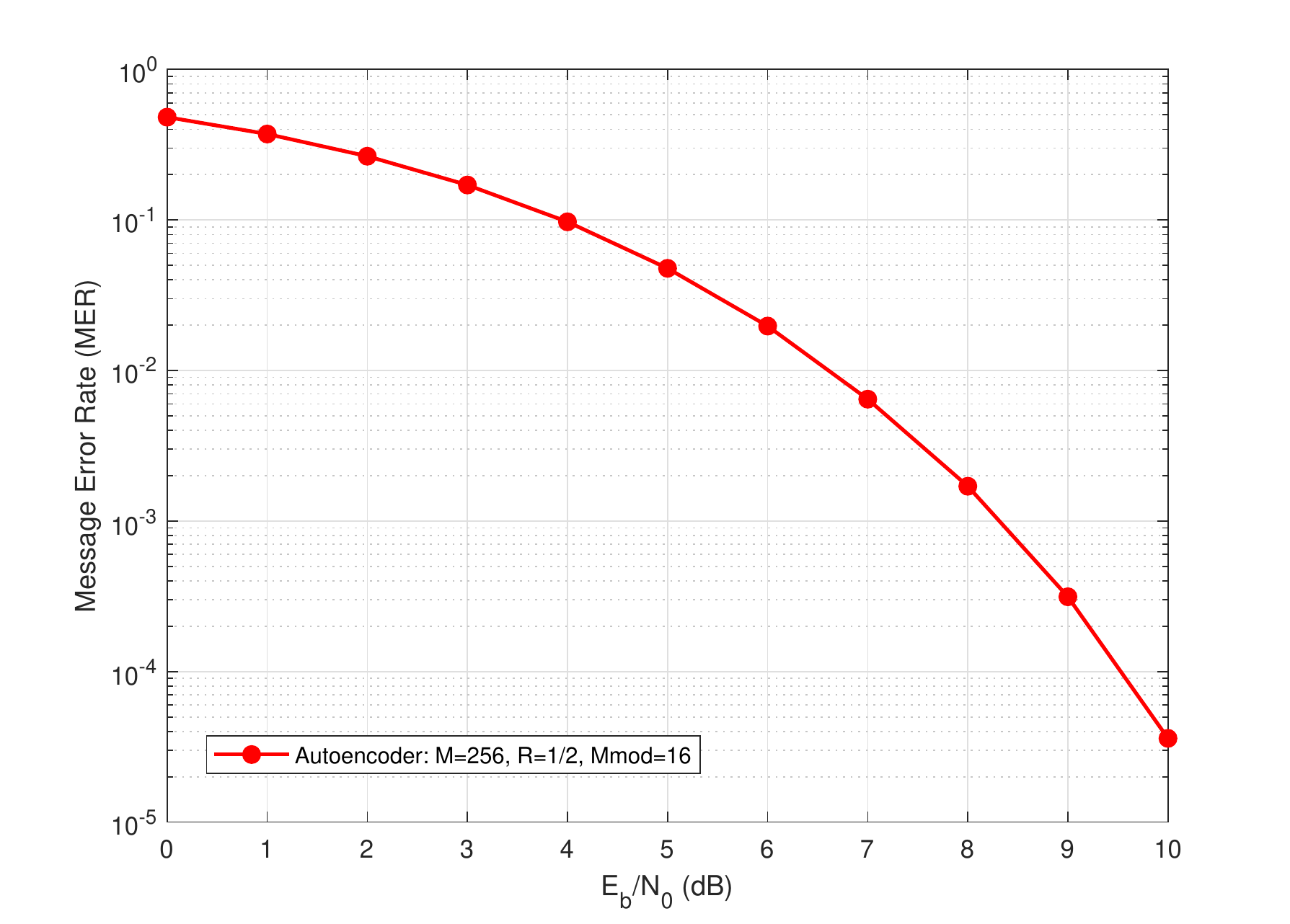}}
\caption{Autoencoder and baseline MER for $R=1/2$ and $M_{mod}=16$.}
\label{fig:mer_M256,Rc05,Mmod16}
\end{figure}

Next we roughly compare the computational complexity of the autoencoder proposed in Section \ref{sec:coded_higher} and the conventional communications system. Here we only consider the processing complexity of the Viterbi decoder when considering the conventional system as it has a higher processing complexity compared to other blocks in the system. Viterbi algorithm with a memory order $m$ has a computational complexity of $4.R.N.2^{m}$ \cite{keith}. Computational complexity in each layer of the autoencoder is given in Table \ref{table:binary_autoencoder_mathematical_operations}. These numbers are further reduced with a parallel processing implementation which is common in neural networks \cite{thesis}. Calculations show that the autoencoder and Viterbi decoder has same range of processing complexity without considering parallel implementation of the autoencoder. Therefore, considering parallel implementation as well, autoencoder has a very low computational complexity compared to the conventional system, given that autoencoder consists of end-to-end processing including both transmitter and receiver sides.

\section{Conclusion}

In this study, we have extended the research done in \cite{oshea2} in autoencoder based end-to-end learning of the physical layer doing a further investigation in order to understand the capabilities of autoencoder based communications systems. Coded BPSK with hard decision CC and equivalent autoencoder implementations have a less than 1 dB gap in BER across the 0-10 dB $E_{b}/N_{0}$ range. Autoencoder is observed to have a closer performance to the baseline for higher code rates. Newly proposed light weight autoencoder is shown competent of learning better communication mechanisms compared to the conventional systems. Simulations show that the autoencoder equivalent of half-rate 16-QAM system has a better BER performance than hard decision CC over the full 0-10 dB $E_{b}/N_{0}$ range and soft decision CC in 0-4 dB $E_{b}/N_{0}$.

\begin{table}[ht]
\caption{Number of mathematical operations in autoencoder model}
\begin{center}
\begin{tabular}{|c||c|c|c|}         
\hline
Layer (output & Multiplications & Additions & Transfer \\     
 dimensions) &  &  & function \\ 
\hline       
Input ($k$)  & -&-&-\\
\hline
Dense-ReLU ($n_{coded}$) & $k.n_{coded}$ & $n_{coded}(k+1)$ & $n_{coded}$ \\
\hline
Dense-ReLU ($2n$) & $n_{coded}.2n$ & $2n(n_{coded}+1)$ & $2n$  \\
\hline
Dense-ReLU ($2n$) & $2n.2n$ & $2n(2n+1)$ & $2n$\\
\hline
Dense-Linear ($2n$) & $2n.2n$ & $2n(2n+1)$ & $2n$\\
\hline
Noise ($2n$) & -&-&- \\
\hline
Dense-ReLU ($2n$) & $2n.2n$ & $2n(2n+1)$ & $2n$\\
\hline
Dense-Linear ($n_{coded}$) & $2n.n_{coded}$ & $n_{coded}(2n+1)$ & $n_{coded}$ \\
\hline
Dense-Sigmoid ($k$) & $n_{coded}.k$ & $k(n_{coded}+1)$ & $k$ \\
\hline
\end{tabular}
\label{table:binary_autoencoder_mathematical_operations}
\end{center}
\end{table}

Autoencoder based end-to-end communications systems, having acceptable BER performance, flexible structure and higher learning capacity, low latency and low processing complexity, show their potential and feasibility as an alternative to conventional communications systems. We have analysed the AWGN channel performance in this study and it should also be extended for other fading channels as well. Also, we have assumed an ideal communications system with perfect timing and carrier-phase and frequency synchronization. Further research needs to be done in order to analyse the performance in non-ideal scenarios. Also, it is essential to investigate the autoencoder performance in comparison to 5G channel codes such as low-density parity-check codes (LDPC) and polar codes.

\bibliography{ref}
\bibliographystyle{IEEEtran}

\vspace{12pt}

\end{document}